\documentclass{aa}  

\usepackage{graphicx}
\usepackage{txfonts}
\usepackage{xcolor}
\usepackage{hyperref}
\hypersetup{colorlinks, linkcolor=blue, citecolor=blue, urlcolor=blue}
\usepackage{amsmath}
\usepackage{ulem}
\usepackage{soul}
\usepackage{enumitem}

\newcommand{\muvec}{\mbox{\boldmath $\mu$}}

\newcommand{\te}{t_{\rm E}}
\newcommand{\thetae}{\theta_{\rm E}}

\newcommand{\pie}{\pi_{\rm E}}
\newcommand{\pien}{\pi_{{\rm E},N}}
\newcommand{\piee}{\pi_{{\rm E},E}}
\newcommand{\dl}{D_{\rm L}}

\definecolor{brown}{rgb}{0.59, 0.29, 0.0}
\definecolor{darkgreen}{rgb}{0.0, 0.42, 0.24}
\definecolor{darkblue}{rgb}{0.01, 0.31, 0.59}
\definecolor{darkblue}{rgb}{0.0, 0.25, 0.42}
\definecolor{blue}{rgb}{0.0,0.0,1.0}
\definecolor{green}{rgb}{0.0,1.0,0.0}

\def\eqalign#1{\null\,\vcenter{\openup\jot
        \ialign{\strut\hfil$\displaystyle{##}$&$
        \displaystyle{{}##}$\hfil \crcr#1\crcr}}\,}

\begin{document} 

\title{KMT-2018-BLG-1988Lb: microlensing super-Earth orbiting a low-mass disk dwarf} 

\author{
     Cheongho Han\inst{\ref{01}} 
\and Andrew Gould\inst{\ref{03},\ref{04}} 
\and Michael~D.~Albrow\inst{\ref{07}} 
\and Sun-Ju~Chung\inst{\ref{08}} 
\and Kyu-Ha~Hwang\inst{\ref{08}} 
\and Youn~Kil~Jung\inst{\ref{08}} 
\and Doeon~Kim\inst{\ref{01}} 
\and Chung-Uk~Lee\inst{\ref{08}} 
\and Shude Mao\inst{\ref{02},\ref{06}}
\and Yoon-Hyun~Ryu\inst{\ref{08}} 
\and In-Gu~Shin\inst{\ref{08}} 
\and Yossi~Shvartzvald,\inst{\ref{09}} 
\and Jennifer~C.~Yee\inst{\ref{10}} 
\and Weicheng Zang\inst{\ref{02}} 
\and Sang-Mok~Cha\inst{\ref{08},\ref{11}} 
\and Dong-Jin~Kim\inst{\ref{08}} 
\and Hyoun-Woo~Kim\inst{\ref{08}} 
\and Seung-Lee~Kim\inst{\ref{08}} 
\and Dong-Joo~Lee\inst{\ref{08}}
\and Yongseok~Lee\inst{\ref{08}} 
\and Byeong-Gon~Park\inst{\ref{08}} 
\and Richard~W.~Pogge\inst{\ref{04}}
\and Chun-Hwey~Kim\inst{\ref{12}}
\\
(The KMTNet Collaboration),\\
}

\institute{
       Department of Physics, Chungbuk National University, Cheongju 28644, Republic of Korea  \\ \email{cheongho@astroph.chungbuk.ac.kr}              \label{01} 
\and   Department of Astronomy, Tsinghua University, Beijing 100084, China                                                                             \label{02} 
\and   Max Planck Institute for Astronomy, K\"onigstuhl 17, D-69117Heidelberg, Germany                                                                 \label{03} 
\and   Department of Astronomy, The Ohio State University, 140 W.  18th Ave., Columbus, OH 43210, USA                                                  \label{04} 
\and   Institute of Natural and Mathematical Sciences, Massey University, Auckland 0745, New Zealand                                                   \label{05} 
\and   National Astronomical Observatories, Chinese Academy of Sciences, Beijing 100101, China                                                         \label{06} 
\and   University of Canterbury, Department of Physics and Astronomy, Private Bag 4800, Christchurch 8020, New Zealand                                 \label{07} 
\and   Korea Astronomy and Space Science Institute, Daejon 34055,Republic of Korea                                                                     \label{08} 
\and   Department of Particle Physics and Astrophysics, WeizmannInstitute of Science, Rehovot 76100, Israel                                            \label{09} 
\and   Center for Astrophysics|Harvard \& Smithsonian 60 Garden St., Cambridge, MA 02138, USA                                                          \label{10}
\and   School of Space Research, Kyung Hee University, Yongin, Kyeonggi 17104, Republic of Korea                                                       \label{11}  
\and   Department of Astronomy \& Space Science, Chungbuk National University, Cheongju 28644, Republic of Korea                                       \label{12}
}
\date{Received ; accepted}

\abstract
{}
{
We reexamine high-magnification microlensing events in the previous data collected by the 
KMTNet survey with the aim of finding planetary signals that were not noticed before.  
In this work, we report the planetary system KMT-2018-BLG-1988L that was found from this 
investigation.
}
{
The planetary signal appears as a deviation with $\lesssim 0.2$~mag from a single-lens 
light curve and lasted for about 6 hours.  The deviation exhibits a pattern of a dip 
surrounded by weak bumps on both sides of the dip.  The analysis of the lensing light 
curve indicates that the signal is produced by a low mass-ratio ($q\sim 4\times 10^{-5}$) 
planetary companion located near the Einstein ring of the host star.
}
{
The mass of the planet, 
$M_{\rm planet}=6.8^{+4.7}_{-3.5}~M_\oplus$ and $5.6^{+3.8}_{-2.8}~M_\oplus$ for the two 
possible solutions, estimated from the Bayesian analysis indicates that the planet is in 
the regime of a super-Earth.  The host of the planet is a disk star with a mass of 
$M_{\rm host} = 0.47^{+0.33}_{-0.25}~M_\odot$ and a distance of $D_{\rm L}= 
4.2^{+1.8}_{-.14}$~kpc.  KMT-2018-BLG-1988Lb is the seventeenth microlensing planet with 
a mass below the upper limit of a super-Earth.  The fact that 14 out of 17 microlensing 
planets with masses $\lesssim 10~M_\oplus$ were detected during the last 5 years since 
the full operation of the KMTNet survey indicates that the KMTNet database is an important 
reservoir of very low-mass planets.  
}
{}


\maketitle

\section{Introduction}\label{sec:one}

Microlensing searches for planets are being conducted by inspecting light curves of more 
than 3000 lensing events, which are annually detected by massive lensing surveys monitoring 
stars lying in the Galactic bulge field.  Planetary signals in lensing light curves appear 
as short-lasting perturbations to the lensing light curves produced by the planet hosts.
Some planetary signals may escape detection for two major reasons.  The first is the short 
duration of a signal relative to the cadence of observations. The duration of a planetary 
signal becomes shorter as the mass ratio $q$ between the planet and host decreases, and thus 
signals of lower-mass planets are more likely to be missed. The second cause arises due to the 
weakness of planetary signals.  A majority of published planetary signals are produced by the 
crossings of source stars over planet-induced caustics, and the signals produced via this 
caustic-crossing channel tend to be strong. However, planetary signals produced via a 
non-caustic-crossing channel tend to be weak, and may escape detection \citep{Zhu2014}. 
Planetary signals can also be weak if lensing events are affected by severe finite-source 
effects or involve faint source stars.

Complete detections of planets including short and weak signals are important for the accurate
estimation of the planet frequency and a solid demographic census of microlensing planetary
systems.  The basis for the statistical assessment of these planet properties is the detection 
efficiency, which is assessed as the ratio of the number of events with detected planets to the 
total number of lensing events, for example, \citet{Gould2010}.  If a planetary 
signal is missed despite its strength being above the detection threshold, the efficiency would 
be underestimated, and this would subsequently lead to the erroneous estimation of planet properties.

From a series of projects conducted during the Covid-19 time, microlensing data collected in 
previous years have been reinvestigated for the purpose of finding missing planets.  
The reinvestigation has been carried out via two approaches.  The first approach is visually 
inspecting lensing light curves with weak or short anomaly features.  From the visual 
reinvestigation of faint-source lensing events in the 2016--2017 season data, \citet{Han2020} 
reported four microlensing planets (KMT-2016-BLG-2364Lb, KMT-2016-BLG-2397Lb, OGLE-2017-BLG-0604Lb, 
and OGLE-2017-BLG-1375Lb), for which the planetary signals had not been noticed due to the weakness 
of the signals caused by the faintness of source stars. From the reexamination of the 2018--2019 
season data, \citet{Han2021a,Han2021b} reported the detections of four planets (KMT-2018-BLG-1025Lb, 
KMT-2018-BLG-1976Lb, KMT-2018-BLG-1996Lb, and OGLE-2019-BLG-0954Lb), for which the planetary signals 
were missed due to their weakness caused by the non-caustic-crossing nature.  The second approach 
was applying an automated algorithm to the previous data to search for buried signals of planets. 
Application of this algorithm to the 2018--2019 prime-field data of the Korea Microlensing Telescope 
Network \citep[KMTNet:][]{Kim2016} survey by \citet{Zang2021b} and \citet{Hwang2021} led to the 
discoveries of six planets (OGLE-2018-BLG-0977Lb, OGLE-2018-BLG-0506Lb, OGLE-2018-BLG-0516Lb, 
OGLE-2019-BLG-1492Lb, and KMT-2019-BLG-0253, OGLE-2019-BLG-1053) with weak and short signals.  
The automated searches for short and weak planetary signatures is planned to be extended to the 
data of the entire KMTNet fields.

In this work, we report the discovery of a super-Earth planet detected from the systematic
reinvestigation of high-magnification lensing events conducted with the aim of finding missing
planets in the 2018 season data obtained by the KMTNet survey.  The planet was not noticed before 
because its signal was not only weak but also short due to its non-caustic-crossing origin and 
the very small planet/host mass ratio.

For the presentation of the planet discovery, we organize the paper according to the following
structure.  In Sect.~\ref{sec:two}, we describe the observation of the lensing event and the 
data used in the analysis.  In Sect.~\ref{sec:three}, we depict the procedure of the lensing light 
curve analysis and present the lens system configuration found from the analysis.  We also mention 
the degeneracy encountered in the interpretation of the lensing event.  In Sect.~\ref{sec:four}, 
we investigate the possibility of measuring higher-order lensing observables. In Sect.~\ref{sec:five}, 
we define the source type by measuring the color and magnitude and estimate the angular Einstein 
radius. In Sect.~\ref{sec:six}, we describe the Bayesian analysis conducted to estimate the physical 
lens parameters. We summarize the result and conclude in Sect.~\ref{sec:seven}.

\section{Observations and data}\label{sec:two}

The planet was detected from the analysis of the lensing event KMT-2018-BLG-1988. The source 
star of the event is located toward the Galactic bulge field with equatorial coordinates 
(R.A., decl.)$_{\rm J2000} = ($17:41:07.81, $-35$:35:40.92), which corresponds to the galactic 
coordinates $(l, b) = (-6^\circ\hskip-2pt .168, -2^\circ\hskip-2pt.686)$.

The lensing event was detected from the post-season inspection of the 2018 KMTNet data using the 
EventFinder algorithm \citep{Kim2018}. The source was located in the KMTNet field BLG37, which was 
covered with a 2.5~hour cadence. This cadence was substantially lower than that of the KMTNet prime 
fields, for which the cadence was 15~minutes. The field corresponds to the BLG610 field of the Optical 
Gravitational Lensing Experiment \citep[OGLE:][]{Udalski2015}, but no OGLE microlensing observation 
was conducted for this field in the 2018 season.

\begin{figure}[t]
\includegraphics[width=\columnwidth]{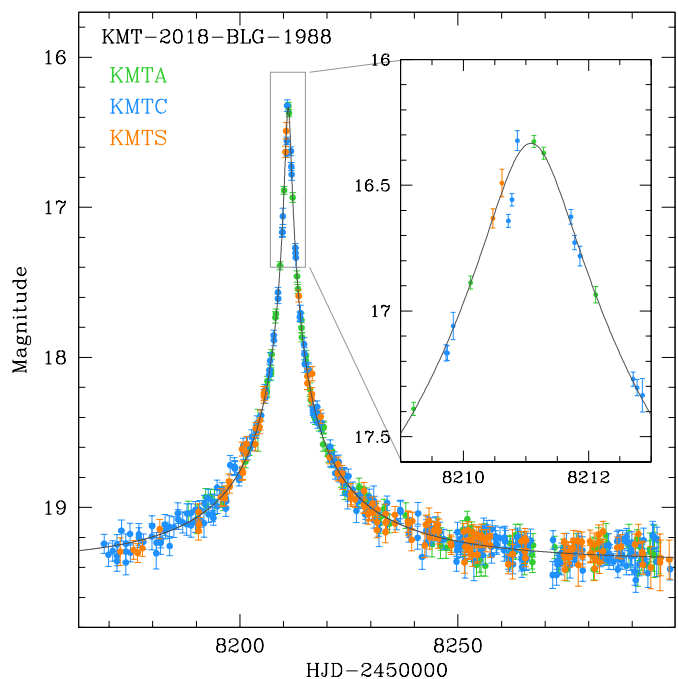}
\caption{
Lensing light curve of KMT-2018-BLG-1988.  The inset shows the zoom-in view of the peak region.
The curve superposed on the data points is the single-lens single-source (1L1S) model. The colors 
of data points are set to match those of the telescopes marked in the legend.
}
\label{fig:one}
\end{figure}

Observations of the event by the KMTNet survey were carried out using the three telescopes that
are located at the Siding Spring Observatory (KMTA) in Australia, the Cerro Tololo Inter-American
Observatory (KMTC) in Chile, and the South African Astronomical Observatory (KMTS) in South
Africa. The telescopes are identical with a 1.6~m aperture, and each  telescope is mounted with a 
camera yielding $2\times 2$~deg$^2$ field of view. Images were taken mainly in the $I$ band, and 
a fraction of $V$-band images were obtained for the source color measurement.

Reduction of the images and photometry of the source were done using the KMTNet pipeline developed 
by \citet{Albrow2009}.  The pipeline is a customized version of the pySIS code established based on 
the difference imaging method \citep{Tomaney1996, Alard1998}.  The error bars of the photometry data 
assessed by the automatized pipeline were readjusted using the method described in \citet{Yee2012} 
in order for them to be consistent with the scatter of data and to make the $\chi^2$ per degree of 
freedom for each data set  unity.  For a subset of KMTC $I$- and $V$-band data, additional photometry 
was done utilizing the pyDIA software \citep{Albrow2017} to construct the color-magnitude diagram (CMD) 
of stars around the source and to measure the reddening- and extinction-corrected (dereddened) source 
color and brightness. We describe the detailed procedure of the source color measurement in 
Sect.~\ref{sec:five}.

\begin{figure}[t]
\includegraphics[width=\columnwidth]{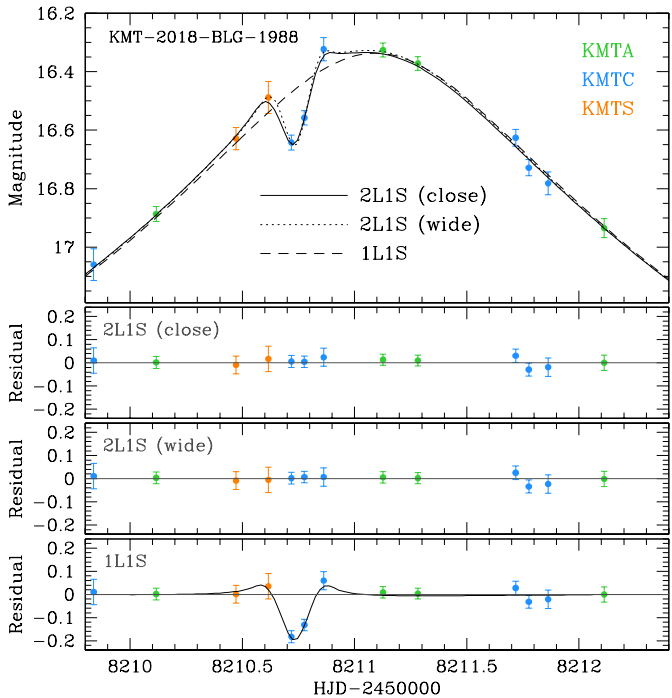}
\caption{
Data around the peak region of the lensing light curve and three tested model curves: 2L1S (close), 
2L1S (wide), and 1L1S.  The three lower panels show the residuals from the individual models. The 
curve drawn in the bottom panel represents the difference between the close 2L1S model and the 1L1S 
model.
}
\label{fig:two}
\end{figure}

Figure~\ref{fig:one} shows the lensing light curve of KMT-2018-BLG-1988 constructed with the 
combined data from the three KMTNet telescopes. At first glance, it appears to show a smooth and 
symmetric form of an event produced by a single-mass lens magnifying a single source (1L1S). 
Drawn over the data points is the 1L1S model curve with the lensing parameters $(u_0, \te)\sim (0.014, 
50~{\rm days})$, where $u_0$ denotes the lens-source separation (scaled to the angular Einstein radius 
$\thetae$) at the time of the closest lens-source approach $t_0$, and $\te$ is the event time scale.  
The full lensing parameters of the 1L1S model is presented in Table~\ref{table:one}.  The event was 
highly magnified with a peak magnification of $A_{\rm peak}\sim 1/u_0\sim 70$. A close look at the 
region around the peak, shown in the inset of Figure~\ref{fig:one} and the top panel of Figure~\ref{fig:two}, 
reveals that there exists a short-term anomaly. The anomaly exhibits deviations with $\lesssim 0.2$~mag 
from the 1L1S model and lasted for about 6~hours.  The anomaly was found from the close inspection of 
the event conducted in the project of reexamining high-magnification events in the previous KMTNet data.  
In this project, high-magnification events are selected as targets for close examination due to their 
high sensitivities to planets \citep{Griest1998}.

\section{Light curve analysis}\label{sec:three}

The signature of the anomaly comes mainly from the two KMTC points acquired at the epochs of 
HJD$^\prime\equiv{\rm HJD} -2450000=8210.719$ and 8210.776, which exhibit deviations from the 1L1S 
model of $\Delta I= -0.182$~mag (compared to the photometric uncertainty of $\sigma_I=0.026$~mag) 
and $-0.132$~mag  ($\sigma_I=0.026$~mag), respectively.  The KMTS point at ${\rm HJD}^\prime=8210.617$
(with $\Delta I=+0.032$~mag) and the KMTC point at ${\rm HJD}^\prime=8210.863$ (with $\Delta I=+0.057$~mag)
also show deviations, although the signals are weaker than the previous two points.
We designate the four points taken at ${\rm HJD}^\prime=8210.617$, $8210.719$, $8210.776$, and $8210.863$
as $t_1$, $t_2$, $t_3$, and $t_4$, respectively.  The residuals from the 1L1S model, presented in the 
bottom panel of Figure~\ref{fig:two}, show that the points at $t_2$ and $t_3$ manifest negative deviations, 
while the data points at $t_1$ and $t_4$ exhibit slight positive deviations, and thus the anomaly appears 
to be a dip surrounded by weak bumps on both sides of the dip.

\begin{figure}[t]
\includegraphics[width=\columnwidth]{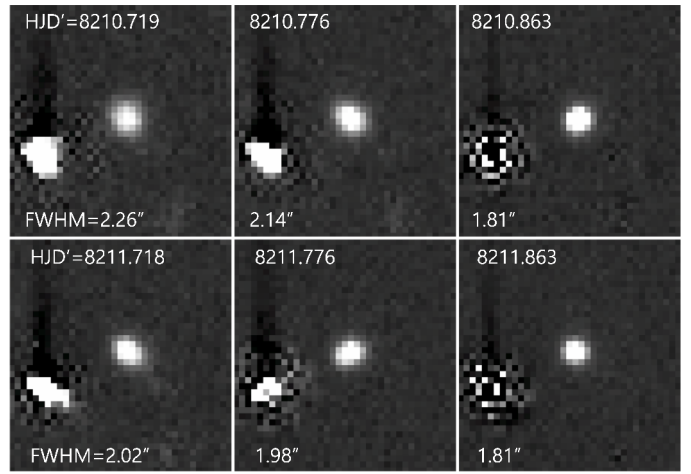}
\caption{
Difference images of the data points taken during (upper three panels) and one day after the 
anomaly (lower panels).  The labels in the upper and lower left side of each image indicate 
the time of the image acquisition and seeing (FWHM).
}
\label{fig:three}
\end{figure}

We checked whether the anomaly were attributed to some kinds of correlated noise, for example, 
fake signals due to correlations with seeing.  For this check, we examined the image quality of 
these data points and compared the quality with that of adjacent data points.  
The upper three panels of Figure~\ref{fig:three} show the difference images obtained by subtracting 
a template image taken before the lensing magnification from the object images of the three anomalous 
KMTC points (at $t_2$, $t_2$, and $t_3$), and they are compared to the other three 
images taken about one day after the anomaly (at HJD$^\prime$=8211.718, 8211.776, and 8211.863). 
In the panels, we insert the seeing (FWHM) values of the individual images.  We find that the 
quality of the images taken during the anomaly is similar to that of the images obtained after the 
anomaly, indicating that there is no noticeable correlations with seeing.  There exists a saturated 
bright star on the lower left side of the source, but we find that it does not affect the photometry 
of the source.  Therefore, we conclude that the anomaly is real.

\begin{table*}[t]
\small
\caption{Lensing parameters of tested models\label{table:one}}
\begin{tabular}{lccccc}
\hline\hline
\multicolumn{1}{c}{Parameter}         &
\multicolumn{1}{c}{1L1S}              &
\multicolumn{2}{c}{2L1S (close)}      &
\multicolumn{2}{c}{2L1S (Wide) }      \\
\multicolumn{1}{c}{}                  &
\multicolumn{1}{c}{}                  &
\multicolumn{1}{c}{Standard}          &
\multicolumn{1}{c}{Higher order}      &
\multicolumn{1}{c}{Standard}          &
\multicolumn{1}{c}{Higher order}      \\
\hline
$\chi^2$                     &  857.2                &  777.3                   &  766.4                    &  777.4                   &  765.8                  \\
$t_0$ (HJD$^\prime$)         &  8211.093 $\pm$ 0.014 &  8211.096 $\pm$ 0.012    &  8211.082 $\pm$ 0.013     &  8211.085 $\pm$ 0.012    &  8211.084 $\pm$ 0.012   \\ 
$u_0$                        &  0.014 $\pm$ 0.001    &  0.014 $\pm$ 0.001       &  0.014 $\pm$ 0.001        &  0.014 $\pm$ 0.001       &  0.0134 $\pm$ 0.001     \\
$\te$ (days)                 &  50.09  $\pm$ 3.26    &  49.38 $\pm$ 3.39        &  50.94 $\pm$ 3.37         &  50.43 $\pm$ 3.23        &  51.19  $\pm$ 3.33      \\
$s$                          &  --                   &  $0.97^{+0.01}_{-0.05}$  &  $0.94^{+0.02}_{-0.05}$   &  $1.01^{+0.09}_{-0.01}$  &  $1.04^{+0.06}_{-0.01}$ \\
$q$ (10$^{-5}$)              &  --                   &  $2.86^{+2.84}_{-0.92}$  &  $4.26^{+4.82}_{-1.30}$   &  $1.74^{+6.94}_{-1.32}$  &  $3.53^{+5.25}_{-0.35}$ \\ 
$\alpha$ (rad)               &  --                   &  2.058 $\pm$ 0.021       &  2.036 $\pm$ 0.023        &  2.036 $\pm$ 0.023       &  2.036 $\pm$ 0.021      \\
$\rho$ (10$^{-3}$)           &  --                   &  $\lesssim 3$            &  $\lesssim 3$             &  $\lesssim 3$            &  $\lesssim 3$           \\
$\pien$                      &  --                   &  --                      &  -0.23 $\pm$ 0.75         &  --                      &  -0.57 $\pm$ 0.76       \\  
$\piee$                      &  --                   &  --                      &  0.19 $\pm$ 0.10          &  --                      &  0.19 $\pm$ 0.10        \\  
$ds/dt$ (yr$^{-1}$)          &  --                   &  --                      &  -0.82 $\pm$ 0.52         &  --                      &  0.18 $\pm$ 0.53        \\  
$d\alpha/dt$ (rad yr$^{-1}$) &  --                   &  --                      &  -0.10 $\pm$ 0.52         &  --                      &  0.03 $\pm$ 0.53        \\  
\hline
\end{tabular}
\tablefoot{ ${\rm HJD}^\prime = {\rm HJD}- 2450000$.  }
\end{table*}

The fact that the anomaly lasted for a short period of time and it occurred near the peak of a 
high-magnification event suggests the possibility that the anomaly was produced by a planetary 
companion. The fact that the anomaly does not exhibit a sharp spike indicates that the source 
did not cross a planet-induced caustic. The anomaly pattern characterized by a dip surrounded by 
shallow bumps suggests that the anomaly was produced by the source star's passage through the 
planet-host axis on the opposite side of the planet.  Example planetary lensing events with 
similar anomaly patterns are found in OGLE-2018-BLG-0677 
\citep{Herrera2020}, KMT-2018-BLG-1976, KMT-2018-BLG-1996 \citep{Han2021a}, KMT-2019-BLG-0253, 
OGLE-2018-BLG-0506, OGLE-2018-BLG-0516, and OGLE-2019-BLG-0149 \citep{Hwang2021}. The possibility 
of a binary-source interpretation \citep{Gaudi1998, Gaudi2004} for the origin of the anomaly is 
excluded because a source companion produces only positive deviations, while the observed anomaly 
exhibits a negative deviation.

Considering the possible planet origin of the anomaly, we conduct a binary-lens (2L1S) modeling
for the observed light curve. A 2L1S lensing light curve is described by 7 parameters: $t_0$, 
$u_0$, $\te$, $s$, $q$, $\alpha$, and $\rho$. The first three $(t_0, u_0, \te)$ are 1L1S parameters 
describing the lens-source approach, and the next three parameters $(s, q, \alpha)$ characterize the 
binary lens, denoting the projected separation (scaled to $\thetae$) and mass ratio between the lens 
components ($M_1$ and $M_2$), and the source trajectory angle defined as the angle between the 
direction of the source motion and $M_1$--$M_2$ axis, respectively. The last parameter (normalized 
source radius), defined as the ratio of the angular source radius $\theta_*$ to the Einstein radius, 
that is, $\rho=\theta_*/\thetae$, is included to account for finite-source effects during caustic 
crossings, although the anomaly is unlikely to be produced by a caustic crossing.

In the 2L1S modeling, we divide the lensing parameters into two groups, in which $(s, q)$ in the
first group are searched for using a grid approach, while the other parameters in the second group
are searched for via a downhill approach. We use the Markov Chain Monte Carlo (MCMC) algorithm for 
the downhill approach. Because a central anomaly can be produced not only by a planetary companion 
lying near the Einstein ring but also by a wide or a close binary companion with a similar mass to 
that of the primary lens \citep{Han2008}, we set the ranges of the grid parameters $s$ and $q$ wide 
enough, $-1.0\leq \log s < 1.0$ and $-5.5\leq \log q < 1.0$, to check the binary origin of the anomaly.  
From this first-round modeling, we identify local solutions in the $\Delta\chi^2$ map on the $s$--$q$ 
plane.  In the second-round modeling, we refine the identified local solutions by releasing all 
parameters as free parameters.

\begin{figure}[t]
\includegraphics[width=\columnwidth]{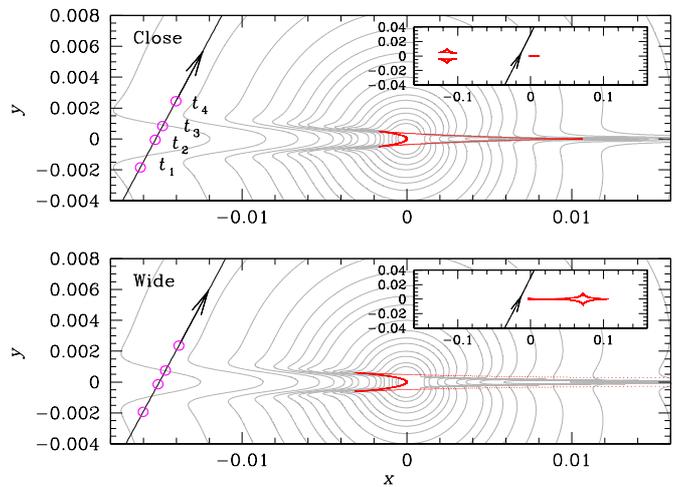}
\caption{
Lens system configurations for the close (upper panel) and wide (lower panel) 2L1S models.  
In each panel, the curve with an arrow is the source trajectory and the cuspy close figures
represent the caustics.  The caustic shape varies depending on time due to the variation of the 
$M_1$--$M_2$ separation induced by the lens orbital motion, and the presented caustic is the one 
corresponding to the time of the anomaly, that is, ${\rm HJD}^\prime\sim 8107.7$.  The grey curves 
around the caustic are the equi-magnification contours.  The four empty circles on the source 
trajectory drawn in magenta color represent the source locations at $t_1=8210.617$, $t_2=8210.719$, 
$t_3=8210.776$, and $t_4=8210.863$.  The size of the circle is arbitrarily set and not scaled to 
the source size.  The inset in each panel shows a wider view including both central and planetary 
caustics. 
}
\label{fig:four}
\end{figure}

In Table~\ref{table:one}, we list the lensing parameters of the solutions found from the 2L1S 
modeling conducted under the assumption of a rectilinear relative lens-source motion (standard 
model). We identify two local solutions with $s_{\rm close}\sim 0.97$ (close solution) and 
$s_{\rm wide}\sim 1.01$ (wide solution), indicating that the companion to the lens is located 
very close to the Einstein ring regardless of the solution.  The estimated lens mass ratio is of 
the order of 10$^{-5}$, indicating that the companion to the lens is a very low-mass planet.  The
model curves of the solutions are drawn over the data points and the residuals from the solutions 
are presented in Figure~\ref{fig:two}.  Both the close and wide 2L1S models explain all the anomalous 
points, improving the fit with $\Delta\chi^2\sim 80$ relative to the 1L1S model.

In Figure~\ref{fig:four}, we present the lens system configurations, which show the source 
trajectory (the curve with an arrow) with respect to the caustic (red concave closed figure), 
of the close (upper panel) and wide (lower panel) solutions.  We note that the presented 
configurations are for the models considering higher-order effects causing non-rectilinear 
lens-source motion to be discussed below, but the configurations of the standard models are 
very similar to the presented ones. As predicted from the pattern and location, the anomaly was 
generated by the passage of the source through the negative anomaly region formed around the 
planet-host axis on the opposite side of the planet. We mark the positions of the source at the 
four epochs of the anomaly: $t_1$, $t_2$, $t_3$, and $t_4$.  The two KMTC data points at $t_2$ 
and $t_3$ correspond to the epochs of the source passage through the negative perturbation region, 
and the other two points at $t_1$ and $t_4$ correspond to the epochs when the source passed the 
positive perturbation regions extending from the back-end cusps of the caustic.

The degeneracy between the close and wide solutions is very severe with $\Delta\chi^2 =0.1$. 
The type of the degeneracy is similar to that identified by \citet{Yee2021}, who found that there 
existed a continuous transition between the two well-known types of degeneracy in planetary lensing 
events: ``close--wide'' \citep{Griest1998, Dominik1999, An2005} and ``inner--outer'' \citep{Gaudi1997} 
degeneracies.  A planetary companion induces two types of caustics, in which one is located close to 
the planet host (central caustic) and the other lies away from the host (planetary caustic).  The 
close-wide degeneracy arises due to the similarity between the two central caustics of the solutions 
with $s<1.0$ and $s>1.0$.  The inner-outer degeneracy arises due to the similarity between the two 
light curves resulting from the source trajectories passing the inner (with respect to the primary) 
and outer region of the planetary caustic.  From the geometry of the source trajectory with respect 
to the central caustic, on one hand, the light curve of KMT-2018-BLG1988 is subject to the close--wide 
degeneracy because the source approached the central caustic.  From the geometry of the source 
trajectory relative to the planetary caustic, on the other hand, the light curve is also subject to 
the inner--outer degeneracy because the source passed the outer (right) and inner (left) region of 
the planetary caustic, as shown in the insets of the individual panels of Figure~\ref{fig:four}.
As a result, the degeneracy is caused by the combination of the two types of degeneracy.

\begin{figure}[t]
\includegraphics[width=\columnwidth]{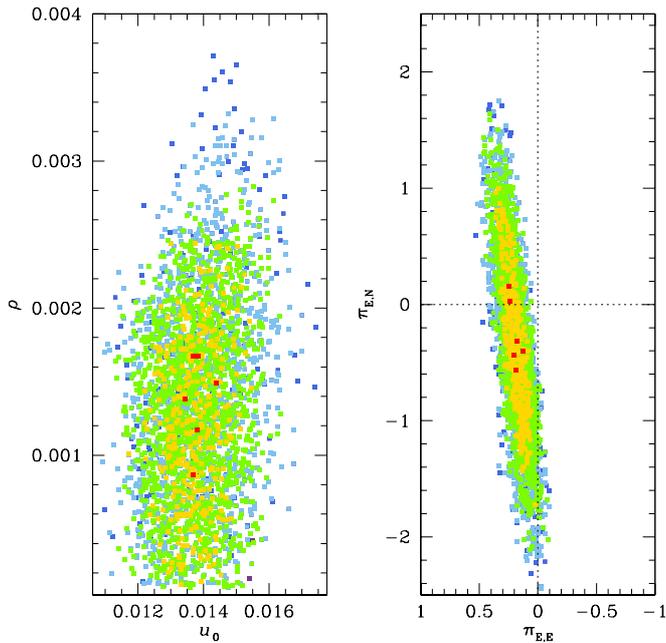}
\caption{
Scatter plots on the $u_0$--$\rho$ (left panel) and $\piee$--$\pien$ (right panel) planes. The 
color coding is set to represent points with $<1\sigma$ (red), $<2\sigma$ (yellow), $<3\sigma$ 
(green), $<4\sigma$ (cyan), and $<5\sigma$ (blue).
}
\label{fig:five}
\end{figure}

\section{Higher-order effects}\label{sec:four}

The physical parameters of the mass $M$ and distance $\dl$ to the lens can be constrained by
measuring the lensing observables including the event time scale $\te$, angular Einstein radius 
$\thetae$, and microlens-parallax $\pie$.  The first two observables are related to $M$ and $\dl$ 
by
\begin{equation}
\te = {\thetae \over \mu};\qquad \thetae = (\kappa M \pi_{\rm rel})^{1/2},
\label{eq1}
\end{equation}
where $\mu$ represents the relative lens-source proper motion, $\kappa =4G/(c^2{\rm AU})$, 
$\pi_{\rm rel}={\rm AU}(D_{\rm L}^{-1}-D_{\rm S}^{-1})$ is the relative lens-source parallax, 
and $D_{\rm S}$ denotes the distance to the source star.  The additional measurement of $\pie$
would enable one to uniquely determine $M$ and $\dl$ by
\begin{equation}
M={\thetae \over \kappa\pie};\qquad
D_{\rm L} = {{\rm AU} \over \pie\thetae + \pi_{\rm S}},
\label{eq2}
\end{equation}
where $\pi_{\rm S}={\rm AU}/D_{\rm S}$ represents the annual parallax of the source star 
\citep{Gould2000}.  The event time scale is measured from the light curve modeling and presented 
in Table~\ref{table:one}. The Einstein radius is estimated by $\thetae=\theta_*/\rho$, where $\rho$ 
is measured from the analysis of the deviations in the lensing light curve caused by finite-source 
effects, and $\theta_*$ is deduced from the source color and brightness. We will discuss in detail 
about the procedure of $\theta_*$ determination in the following section. The microlens parallax is 
measured from the deviation of the light curve caused by the non-rectilinear lens-source motion 
induced by the orbital motion of Earth around the Sun \citep{Gould1992}. For the measurement $\pie$, 
we conduct an additional modeling considering the microlens-parallax effect. In this modeling, we 
also consider the orbital motion of the lens \citep{Dominik1998}, which is known to produce similar 
deviations to those induced by the parallax effect \citep{Batista2011}. The inclusion of the higher-order 
effects in modeling requires one to include two extra pairs of lensing parameters: $(\piee, \pien)$ and 
$(ds/dt, d\alpha/dt)$. The first pair denote the north and east components of the microlens parallax 
vector $\pie=(\pi_{\rm rel}/ \thetae)(\muvec/\mu)$, respectively, and the other pair represent the 
annual change rates of the binary separation and source trajectory angle, respectively.

The lensing parameters obtained from the modeling considering the higher-order effects (higher-order 
2L1S model) are listed in Table~\ref{table:one}. We find that the consideration of the higher-order 
effects improves the fit by $\Delta\chi^2\sim 11$ with respect to the standard models. It is found 
that finite-source effects cannot be firmly detected because the source did not cross the caustic.  
Nevertheless, the upper limit of $\rho$ can be placed. The left panel of Figure~\ref{fig:five} 
shows the distribution of points in the MCMC chain (scatter plot) on the $u_0$--$\rho$ parameter 
plane. We set a conservative upper limit as $\rho_{\rm max}=0.003$. The right panel of 
Figure~\ref{fig:five} shows the scatter plot on the $\piee$--$\pien$ plane. 
It shows that the uncertainties of the parallax parameters are big, especially the north component
$\pien$. Similarly, the uncertainties of the orbital parameters are considerable.

\section{Source star}\label{sec:five}

For the estimation of the angular Einstein radius, we specify the source type by measuring its 
dereddened color and brightness, $(V-I, I)_0$. In order to measure $(V-I, I)_0$ from the uncalibrated 
values, $(V-I, I)$, in the instrumental color-magnitude diagram (CMD), we use the \citet{Yoo2004} 
method, in which the centroid of red giant clump (RGC) is used for calibration.

\begin{figure}[t]
\includegraphics[width=\columnwidth]{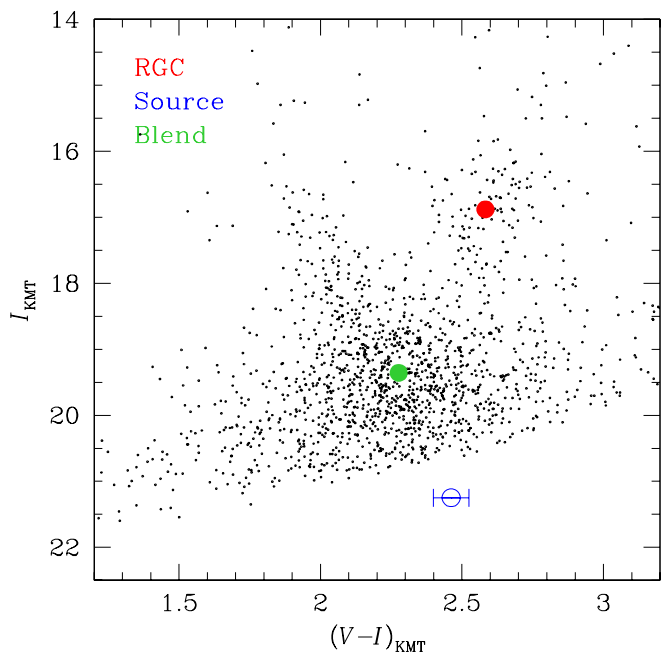}
\caption{
Source location (blue empty dot) with respect to the centroid of red giant clump (RGC, filled red 
dot) in the color-magnitude diagram constructed using the pyDIA photometry of the KMTC data. Also 
marked is the location of the blend (green filled dot).
}
\label{fig:six}
\end{figure}

Figure~\ref{fig:six} shows the source location (empty circle with error bars) with respect to 
the RGC centroid (red filled dot) in the instrumental CMD of stars around the source constructed 
using the pyDIA photometry of the KMTC $I$- and $V$-band data.  Also marked is the location of 
the blend (green filled dot). The measured instrumental colors and magnitudes of the source and 
RGC centroid are $(V-I, I)=(2.462\pm 0.063, 21.248\pm 0.002)$ and $(V-I,I)_{\rm RGC}=(2.583, 16.882)$, 
respectively. From the offsets in color and magnitude between the source and RGC centroid, 
$\Delta (V-I, I)$, together with the known dereddened values of the RGC centroid $(V-I,I)_{\rm RGC,0}
=(1.060, 14.619)$ from \citet{Bensby2013} and \citet{Nataf2013}, we estimate the dereddened color 
and magnitude of the source as
\begin{equation}
\eqalign{
(V-I, I)_0 =  & (V-I, I)_{\rm RGC,0} + \Delta(V-I, I) \cr 
           =  & (0.939 \pm 0.063, 18.985 \pm 0.002),  \cr
}
\label{eq3}
\end{equation}
indicating that the spectral type of the source is K2V.

The angular source radius is estimated by first converting the measured $V-I$ color into $V-K$ 
color using the color-color relation of \citet{Bessell1988}, and then deduce $\theta_*$ from the 
$(V-K)$--$\theta_*$ relation of \citet{Kervella2004}. From this procedure, it is estimated that 
the source has an angular radius of
\begin{equation}
\theta_* = 0.650 \pm 0.061~\mu{\rm as}. 
\label{eq4}
\end{equation}
Combined with the measured event time scale and the upper limit of the normalized source radius, 
the lower limits of the Einstein radius and the relative lens-source proper motion are set as
\begin{equation}
\theta_{\rm E,min} ={\theta_*\over \rho_{\rm max}} = 0.22~{\rm mas},
\label{eq5}
\end{equation}
and
\begin{equation}
\mu_{\rm min} ={\theta_{\rm E, min} \over \te} = 1.62~{\rm mas}~{\rm yr}^{-1}, 
\label{eq6}
\end{equation}
respectively.

We check whether a significant fraction of the blended light comes from the lens. The flux from 
a lens contributes to the blended flux, and thus if the lens is bright enough, the lens can be 
additionally constrained by analyzing the blended flux, for example, the planetary events 
OGLE-2018-BLG-1269 \citep{Jung2020a} and OGLE-2018-BLG-0740 \citep{Han2019}. For this check, we 
measure the astrometric offset between the baseline object and the source. The position of the 
source is measured from the difference images obtained during lensing magnifications. The measured 
offsets in the east and north directions are $\Delta\theta (E, N) = (37.2\pm 5.6, 370.0\pm 4.8)$~mas. The 
offset is well beyond the measurement error, indicating that the blended flux is likely to come from 
an adjacent star unrelated to lensing.  Considering that the blend lies on the main-sequence branch 
of disk stars, it is unlikely that the blend is a binary companion to the source which lies in the 
bulge.  This is also supported by the fact that the lens brightness expected from the mass and 
distance, to be estimated in the next section, is much fainter than the brightness of the blend.

\begin{figure}[t]
\includegraphics[width=\columnwidth]{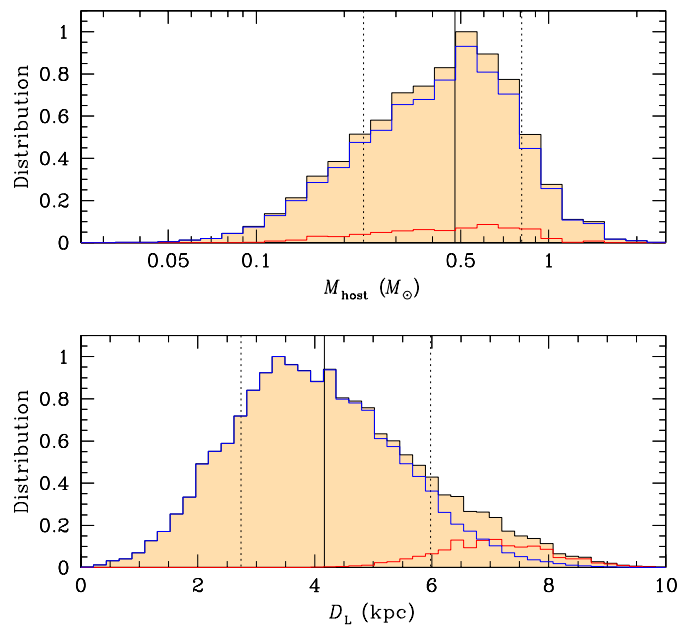}
\caption{
Bayesian posterior distributions of the planet host mass and distance to the lens. In each panel, 
the blue and red curves indicate the contributions by the disk and bulge lens populations. The 
solid vertical line represents the median of the distribution and the two dotted lines denote 
1$\sigma$ range of the distribution.
}
\label{fig:seven}
\end{figure}

\begin{table}[t]
\small
\caption{Physical lens parameters\label{table:two}}
\begin{tabular*}{\columnwidth}{@{\extracolsep{\fill}}lcccc}
\hline\hline
\multicolumn{1}{c}{Parameter}    &
\multicolumn{1}{c}{Close}        &
\multicolumn{1}{c}{Wide }        \\
\hline
$M_{\rm host}$ ($M_\odot$)      &  $0.48^{+0.33}_{-0.25}$  &  $0.47^{+0.33}_{-0.24}$ \\
$M_{\rm planet}$ ($M_\oplus$)  &  $6.78^{+4.68}_{-3.47}$  &  $5.56^{+3.84}_{-2.84}$ \\ 
$\dl$ (kpc)                     &  $4.17^{+1.81}_{-1.43}$  &  $4.15^{+1.79}_{-1.42}$ \\
$a_\perp$ (AU)                  &  $2.83^{+1.23}_{-0.97}$  &  $3.09^{+1.33}_{-1.05}$ \\
\hline
\end{tabular*}
\end{table}

\begin{table*}[t]
\small
\caption{Microlensing planets with masses less than $10~M_\oplus$ \label{table:three}}
\begin{tabular}{llll}
\hline\hline
\multicolumn{1}{c}{Planet}                             &
\multicolumn{1}{c}{$M_{\rm planet}$ ($M_\oplus$)}     &
\multicolumn{1}{c}{$M_{\rm host}$ ($M_\odot$)}         &
\multicolumn{1}{c}{Reference and comment}              \\
\hline
OGLE-2005-BLG-390Lb     &  $5.5^{+5.5}_{-2.7} $                   &  $0.22^{+0.21}_{-0.11}$                &  \citet{Beaulieu2006}                         \\
MOA-2007-BLG-192Lb      &  $3.3^{+4.9}_{-1.6} $                   &  $0.06^{+0.028}_{-0.021}$              &  \citet{Bennett2008}                          \\ 
OGLE-2013-BLG-0341Lb    &  $1.66\pm 0.18$ (solution 1)            &  $0.11\pm 0.01$ (solution 1)           &  \citet{Gould2014}                            \\
                        &  $2.32\pm 0.27$ (solution 2)            &  $0.15\pm 0.01$ (solution 2)           &                                               \\
OGLE-2016-BLG-1195Lb    &  $5.10^{+4.96}_{-2.85}$                 &  $0.37^{+0.38}_{-0.21}$                &  \citet{Bond2017}, \citet{Shvartzvald2017}    \\
OGLE-2016-BLG-1928L     &  $\sim 0.1$--$1.0 $                     &         --                             &  \citet{Mroz2020}, free-floating planet       \\
OGLE-2017-BLG-1434Lb    &  $4.4\pm0.5$                            &  $0.23\pm 0.03$                        &  \citet{Udalski2018}, \citet{Blackman2021}    \\ 
KMT-2018-BLG-0029Lb     &  $7.59^{+0.75}_{-0.69}$                 &  $1.14^{+0.10}_{-0.12}$                &  \citet{Gould2020}                            \\
OGLE-2018-BLG-0532Lb    &  $6.29^{+0.91}_{-0.89}$ (solution 1)    &  $0.20^{+0.02}_{-0.02}$ (solution 1)   &  \citet{Ryu2020}                              \\
                        &  $6.55^{+0.91}_{-0.81}$ (solution 2)    &  $0.20^{+0.02}_{-0.02}$ (solution 2)   &  \citet{Ryu2020}                              \\
OGLE-2018-BLG-0677Lb    &  $3.96^{+5.88}_{-2.66}$                 &  $0.12^{+0.14}_{-0.08}$                &  \citet{Herrera2020}                          \\  
OGLE-2018-BLG-0977Lb    &  $\sim 6.4^{+5.2}_{-3.7} $              &  $0.47^{+0.38}_{-0.27}$                &  \citet{Hwang2021}                            \\  
KMT-2018-BLG-1025Lb     &  $6.06^{+8.20}_{-3.32} $  (solution 1)  &  $0.21^{+0.30}_{-0.12} $ (solution 1)  &  \citet{Han2021a}                             \\  
                        &  $4.44^{+6.80}_{-2.41} $  (solution 2)  &  $0.08^{+0.13}_{-0.04} $ (solution 2)  &                                               \\  
KMT-2018-BLG-1988Lb     &  $6.8^{+4.7}_{-3.5}$                    &  $0.47^{+0.33}_{-0.25}$                &  This work                                    \\ 
                        &  $5.6^{+3.8}_{-2.8}$                    &  --                                    &                                               \\
KMT-2019-BLG-0253Lb     &  $9.2^{+5.0}_{-4.1}$                    &  $0.70^{+0.34}_{-0.31} $               &  \citet{Hwang2021}                            \\
KMT-2019-BLG-0842Lb     &  $10.3\pm 5.5$                          &  $0.76\pm 0.40$                        &  \citet{Jung2020b}                            \\
KMT-2019-BLG-0960Lb     &  1.4 -- 3.1                             &  0.3 -- 0.6                            &  \citet{Yee2021}                              \\
OGLE-2019-BLG-1053Lb    &  $2.48^{+1.19}_{-0.98}$                 &  $0.61^{+0.29}_{-0.24}$                &  \citet{Zang2021b}                            \\  
KMT-2020-BLG-0414LAb    &  $.3\pm 0.1$                            &  $1.0\pm 0.3$                          &  \citet{Zang2021a}, planet in a binary        \\  
\hline
\end{tabular}
\end{table*}

\section{Physical parameters of planetary system}\label{sec:six}

We estimate the physical parameters of the planetary system based on the measured lensing observables. 
Among the lensing observables of $(\te, \thetae, \pie)$, the event time scale is relatively well 
constrained, but the uncertainty of the microlens parallax is substantial, and only the upper limit 
of the Einstein radius is constrained, making it difficult to uniquely determine $M$ and $\dl$ 
using the relation in Equation~(\ref{eq2}). We, therefore, estimate the physical lens parameters by 
conducting a Bayesian analysis using the constraints from the observables together with a prior 
Galactic model.

In the Bayesian analysis, we first conduct a Monte Carlo simulation to produce a large number 
($2\times 10^6$) of artificial lensing events, for which the locations of the lens and source, 
lens-source transverse speeds, $v_\perp$, and lens masses are derived from the Galactic model. The 
Galactic model is constructed by adopting the \citet{Han2003} physical distribution, \citet{Han1995} 
dynamical distribution, and \citet{Zhang2020} mass function of Galactic objects.  The further 
description about the Galactic model is found in \citet{Jung2021}. For the individual simulated 
events, we compute lensing observables by $\te =\dl\thetae/v_\perp$, $\thetae=(\kappa 
M\pi_{\rm rel})^{1/2}$, and $\pie=\pi_{\rm rel}/\thetae$, and then construct the posterior 
distributions of the lens mass and distance for the events with observables lying within the 
ranges of the observed values.

Figure~\ref{fig:seven} shows the posterior distributions of the planet host mass and distance to the 
planetary system. In Table~\ref{table:two}, we summarize the masses of the host, $M_{\rm host}$, 
and planet, $M_{\rm planet}$, distance, and projected host-planet separation, $a_\perp$, estimated 
based on the lensing parameters of the close and wide solutions.  For each parameter, the median of 
the Bayesian distribution is taken as a represent value, and the lower and upper limits are set as 
the 16\% and 84\% of the distribution, respectively. The estimated mass, $M_{\rm planet}=
6.8^{+4.7}_{-3.5}~M_\oplus$ for the close solution and $5.6^{+3.8}_{-2.8}~M_\oplus$ for the wide 
solution, indicates that the planet is in the regime of a super-Earth, which is defined as a planet 
with a mass higher than the mass of Earth, but substantially lower than those of the Solar System's 
ice giants, that is, Uranus and Neptune \citep{Valencia2007}. The mass of the planet host, 
$M_{\rm host} = 0.47^{+0.33}_{-0.25}~M_\odot$, indicates that the host is a late K or an early 
M dwarf. The estimated distance to the lens, $D_{\rm L}= 4.2^{+1.8}_{-.14}$~kpc, suggests that the 
lens is likely to lie in the disk. According to the Bayesian estimation of the disk (blue curve in 
Figure~\ref{fig:seven}) and bulge (red curve) contributions, the probability of a disk lens is 
$\sim 82\%$.  We note that the uncertainties of the estimated physical lens parameters are considerable 
due to the combination of the statistical nature of the Bayesian analysis and the weak constraints of 
lensing observables.  Furthermore, the estimated parameters can vary depending on adopted models 
\citep{Yang2021}.

The planetary system KMT-2018-BLG-1988L demonstrates the importance of high-cadence microlensing 
surveys in finding low-mass planets.  By setting the upper mass limit of a super-Earth as $\sim 10~M_\oplus$, 
there exist 17 microlensing planetary systems with planet masses less than this upper limit, including 
the system reported in this work. In Table~\ref{table:three}, we list these planets discovered so far 
from the lensing surveys that have been conducted for almost three decades.  In the table, we list the 
masses of planets and hosts together with the references and brief comments for notable systems.  We 
note that the list is arranged according to the chronological order of the lensing event discovery not 
by the times of the planet discovery.  Among these discovered planets, 14 planets have been found since 
2016, when the KMTNet survey conducted its full operation, indicating that the KMTNet database is an 
important reservoir of low-mass microlensing planets.

\section{Conclusion}\label{sec:seven}

We reported the super-Earth planet KMT-2018-BLG-1988L detected from the reinvestigation of 
high-magnification events in the previous microlensing data collected by the KMTNet survey.  The 
planetary signal appeared as a deviation with $\lesssim 0.2$~mag from a 1L1S curve and lasted for 
about 6 hours.  The analysis of the lensing light curve indicated that the signal was produced by 
a low mass-ratio planetary companion located at around the Einstein ring of the host star.  The mass 
of the planet, $M_{\rm planet}=6.8^{+4.7}_{-3.5}~M_\oplus$ and $5.6^{+3.8}_{-2.8}~M_\oplus$ for the 
two possible solutions, estimated from the Bayesian analysis indicated that the planet was in the 
regime of a super-Earth.  The planet belonged to a disk star with a mass of $M_{\rm host} = 
0.47^{+0.33}_{-0.25}~M_\odot$ located at a distance of $D_{\rm L}= 4.2^{+1.8}_{-.14}$~kpc.  
KMT-2018-BLG-1988L is the seventeenth microlensing planet with a mass below the upper limit of a 
super-Earth.  The fact that 14 out of 17 planets with masses $\lesssim 10~M_\oplus$ were detected 
during the last 5 years since the full operation of the KMTNet survey indicates that the KMTNet 
database is an important reservoir of very low-mass planets.

\begin{acknowledgements}
Work by C.H. was supported by the grants  of National Research Foundation of Korea 
(2020R1A4A2002885 and 2019R1A2C2085965).
This research has made use of the KMTNet system operated by the Korea Astronomy and Space 
Science Institute (KASI) and the data were obtained at three host sites of CTIO in Chile, 
SAAO in South Africa, and SSO in Australia.
\end{acknowledgements}

\end{document}